\documentclass[conference]{IEEEtran}

\usepackage{amsmath, graphicx, amssymb, amsfonts, amsthm, subfigure, arydshln, setspace}

\newcommand {\Graph} {{\mathcal G}}
\newcommand {\Edges} {{\cal E}}

\newcommand {\Nodes} {{\mathcal V}}
\newcommand {\F} {\mathbb F}
\newcommand {\DisPath} {\mathcal P}
\newcommand {\Receiver} {\mathcal {R}}
\newcommand {\Polynomial} {poly}
\newcommand {\SourceNum}{s}
\newcommand {\NetSize} {|\mathcal {E}|}
\newcommand {\Exponential}{exp}

\newtheorem{theorem}{Theorem}

\newtheorem{lemma}{Lemma}
\newtheorem{proposition}{Proposition}

\begin{document}

\title{Multiple Access Network Information-flow And Correction codes$^*$}

\author{
\authorblockA{$\text{Hongyi Yao}^1$, $\text{Theodoros K. Dikaliotis}^2$, $\text{Sidharth Jaggi}^3$, $\text{Tracey Ho}^2$\\
$^1$Tsinghua University $\hspace{1mm}^2$California Institute of
Technology $\hspace{1mm}^3$Chinese University of Hong Kong\\
{$^1$yaohongyi03@gmail.com $^2$\{tdikal, tho\}@caltech.edu
$^3$jaggi@ie.cuhk.edu.hk}}}

\maketitle

\begin{abstract}
The network communication scenario where one or more receivers
request all the information transmitted by different sources is
considered. We introduce  distributed polynomial-time network codes
in the presence of {\it malicious nodes}. Our codes can achieve any
point inside the rate region of multiple-source multicast
transmission scenarios both in the cases of coherent and
non-coherent network coding. For both cases the encoding and
decoding algorithm runs in
$\Polynomial(\NetSize)\Exponential(\SourceNum)$ time, where
$\Polynomial(\NetSize)$ is a polynomial function of the number of
edges $\NetSize$ in the network and $\Exponential(\SourceNum)$ is an
exponential function of the number of sources $\SourceNum$. Our
codes are fully distributed and different sources require no
knowledge of the data transmitted by their peers. Our codes are
``end-to-end'', that is, all nodes apart from the sources and the
receivers are oblivious to the adversaries present in the network
and simply implement random linear network coding.
\let\thefootnote\relax\footnotetext{$^*$ In other words, MANIAC codes.}

\begin{keywords}
polynomial-time codes, error-correction, double extended field, Gabidulin codes
\end{keywords}
\end{abstract}

\section{Introduction}

Information dissemination can be optimized with the use of network
coding since it maximizes the network throughput in multicast
transmission scenarios~\cite{ahlswede00network}. At the same time
network coding is highly vulnerable to malicious attacks from rogue
users. The presence of even a small number of adversarial nodes can
contaminate the majority of packets in a network, preventing
receivers from decoding.

The work of Cai-Yeung~\cite{RaymondNECC2002} first studied the
network error-correction problem in the single source scenario, and
their scheme requires high (exponential in the network size) design
complexity. Further works by~\cite{jaggi07resilient}
and~\cite{RankMetricRandCodes} provided network error-correcting
codes with design and implementation complexity that is low ({\it
i.e.}, polynomial in size of the network parameters). The design of
such robust network codes with ``active nodes" ({\it i.e.} internal
nodes using cryptographic schemes to detect packets modified by
computationally bounded adversaries) has also been considered in the
cryptographic setting (see for
instance~\cite{gkantsidis06:cooperative, charles06signaturesfor}).

We consider the design of multisource network error-correcting codes
that are resilient against worst-case network errors, {\it i.e.},
against errors injected by computationally unbounded adversaries.
Na\"ive implementations of single source network error-correcting
codes fail since such codes require the source to judiciously insert
redundancy into the transmitted codeword; however, in the
distributed source case this cannot be done. The work
in~\cite{Svit_rate_regions} gave the capacity region for the
multisource network error correction problem, but the achievability
proof used codes with high decoding complexity.

The current paper gives the first construction of efficient
decodable error-correction codes. For both coherent (when the
network transform is known {\it a priori} to the receiver(s)) and
non-coherent (when no such information is known {\it a priori} to
the receiver(s)) cases our codes achieve the optimal rate-region
demonstrated in~\cite{Svit_rate_regions} and have implementation
complexity that is polynomial in the size of the network.
Furthermore our codes are fully distributed in the sense that
different sources require no knowledge of the data transmitted by
their peers and end-to-end, {\it i.e.}, all nodes are oblivious to
the adversaries present in the network and simply implement random
linear network coding~\cite{RandCode0}. A remaining bottleneck is
that the computational complexity of our codes increases
exponentially with the number of sources. Thus the design of
efficient schemes for a large number of sources is still open.

The remainder of this paper is organized as follows: In
Section~\ref{Sec:Pre} we formulate the problem and introduce the
mathematical preliminaries. In Section~\ref{The_Coherent_Case} we
provide a code construction for the coherent case, {\it i.e.}, the
receiver(s) knows the linear transform from each source induced by
random linear network code. In Section~\ref{The_Non_Coherent_Case}
we construct codes for the non-coherent case, where the receiver(s)
has no information on the network transforms.

\section{Preliminaries}
\label{Sec:Pre} \vspace{-2.8mm}
\subsection{Model} \label{The_Model}

We consider a delay-free network $\Graph=(\Nodes,\Edges)$ where
$\Nodes$ is the set of nodes and $\Edges$ is the set of edges. The
capacity of each edge is normalized to be one symbol of $\F_p$ per
unit time.  Edges with non-unit capacity are modeled as parallel
edges.

For notational convenience we restrict  ourselves to the analysis of
the situation where there are only two sources
$\mathcal{S}_1,\mathcal{S}_2\in \Nodes$ transmitting information to
one receiver $\Receiver\in \Nodes$, since the extension of our
results to more sources and receivers is straightforward. The
minimum cut capacity from source $\mathcal{S}_i$ to $\mathcal{R}$ is
denoted by $C_i$ for $i\in[1,2]$, and the minimum cut capacity from
both sources to the receiver is equal to $C$.

Within the network there is a hidden adversary trying to interfere
with the transmission of information by observing all the
transmissions in the network and injecting its own packets in any
$z$ links\footnote{Note that since each transmitted symbol in the
network is from a finite field, modifying symbol $x$ to symbol $y$
is equivalent to injecting/adding symbol $y-x$ into $x$.}, that may
be chosen as a function of his complete knowledge of the network,
the message, and the communication scheme.

\begin{figure}[htp]
  \begin{center}
    \subfigure[Example Network]{\label{ToyExample-a}\includegraphics[height=30mm,width=30mm]{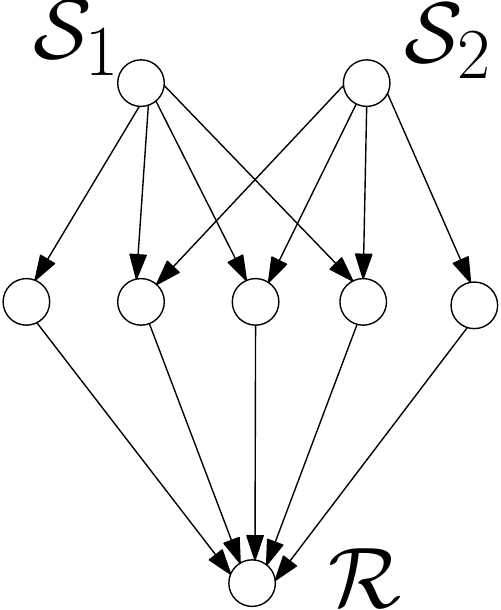}}
    \quad\qquad\subfigure[Rate Region]{\label{ToyExample-b}\includegraphics[height=30mm,width=30mm]{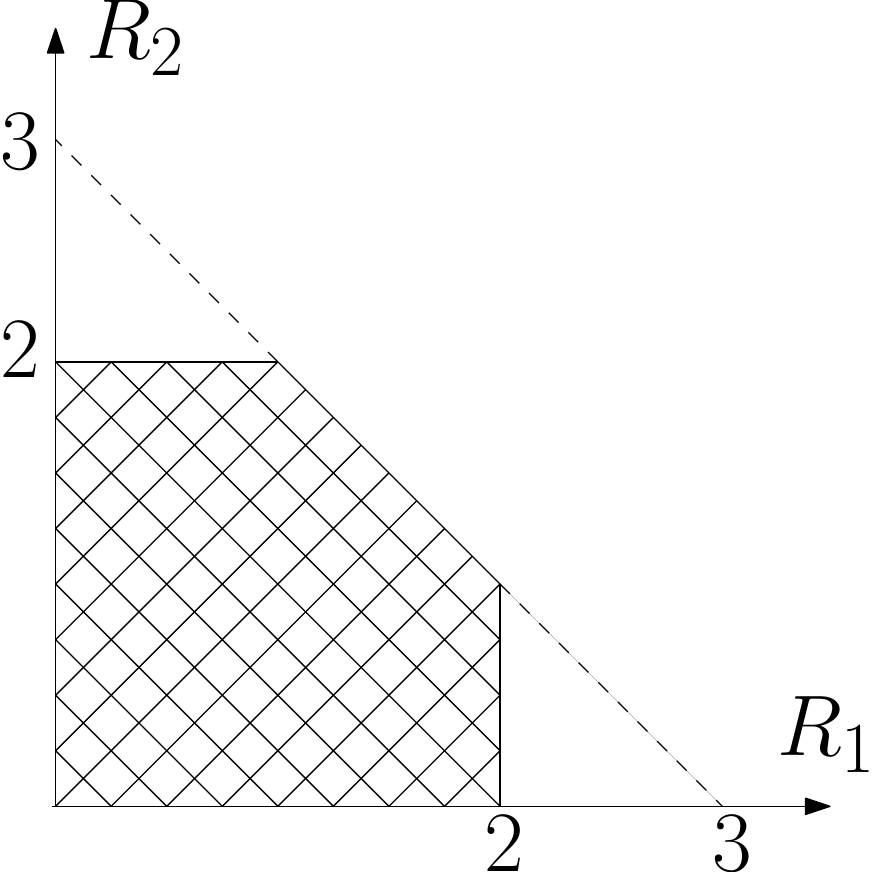}} \\
  \end{center}
  \caption{An example network with two sources. The Network in Figure~\ref{ToyExample-a} has $C_1=C_2=4$, $C=5$
  and the adversary can inject $z=1$ error packet. The achievable rate region is shown
 in the dark region of Figure~\ref{ToyExample-b}.}
  \label{ToyExample}
\end{figure}

The sources on the other hand know nothing about each other's
transmitted information and the links compromised by the adversary.
Their goal is to add redundancy into their transmitted packets so
that they can achieve any rate-tuple $(R_1, R_2)$ such that $R_1\leq
C_1-2z$, $R_2\leq C_2-2z$, and $R_1+R_2\leq C-2z$ (this is the rate
region of the multi-source multicast problem proved
in~\cite{Svit_rate_regions}). An example network and its rate region
is shown in Figure~\ref{ToyExample}.

To simplify the discussion we show the code construction for
rate-tuple $(R_1,R_2)$ satisfying $R_1\leq C_1-2z$, $R_2\leq
C_2-2z$, $R_1+R_2+2z=C$ and exactly $C$ edges reach the receiver
$\mathcal R$ (if more do, redundant information can be discarded).

\subsection{Random linear network coding}
\label{subsec:random-linear-netcod} In this paper, we consider the following well-known distributed random linear coding scheme~\cite{RandCode0}.

{\it Sources:} The source $\mathcal {S}_i$ arranges the data into a
$C_i \times \ell$ {\it message matrix} $M_i$ over $\F_p$ (here the
{\it packet-length} $\ell$ is a network design parameter). For
$i\in[1,2]$ source $\mathcal {S}_i$ then takes independent and
uniformly random linear combinations over $\F_p$ of the rows of
$M_i$ to generate respectively the packets transmitted on each
outgoing edge.

{\it Network nodes:} Each internal node similarly takes uniformly
random linear combinations of the packets on incoming edges to
generate packets transmitted on outgoing edges.

{\it Receiver:} The receiver $\Receiver$ constructs the $C \times
\ell$ matrix $Y$ over $\F_p$ by treating the received packets as
consecutive length-$\ell$ row vectors of $Y$ (recall that exactly
$C$ edges reach $\Receiver$). In the case that no error happen in
the network, the network's internal linear operations induce linear
transforms between $M_i$ and $Y$ as
\begin{equation}
Y =T_1 M_1+ T_2M_2, \label{eq:net_tran}
\end{equation}
where $T_i$ is the overall transform matrix from $\mathcal {S}_i$ to
$\Receiver$.

\subsection{Finite field extension}
\label{sec:Field_extension} In the analysis below denote by
$\F_p\hspace{-1.1mm}^{m\times n}$ the set of all $m\times n$
matrices with elements from $\F_p$. The identity matrix with
dimension $m\times m$ is denoted by $I_m$, and the zero matrix of
any dimensions is denoted by $O$ .The dimension of the zero matrix
will be clear from the context stated. For the clarity of notation
vectors are in bold-face ({\it e.g.} ${\bf A}$).

Before we continue to the analysis of the encoding and decoding
process it is useful to introduce some concepts from the theory of
finite field. Every finite field $\F_p$, where $p$ is a prime or a
power of a prime, can be algebraically extended\footnote{Let
$\F_p[x]$ be the set of all polynomials over $\F_p$ and $f(x)\in
\F_p[x]$ be an irreducible polynomial of degree $n$. Then
$\F_p[x]/f(x)$ defines an algebraic extension field $\F_{p^n}$ by a
homomorphism mapping~\cite{Algebra_Martin}.}~\cite{Algebra_Martin}
to a larger finite field $\F_{q}$, where $q=p^n$ for any positive
integer $n$. Note that $\F_{q}$ includes $\F_{p}$ as a subfield thus
any matrix $A\in \F_p\hspace{-1.1mm}^{m\times\ell}$ is also a matrix
in $\F_{q}\hspace{-1.1mm}^{m\times \ell}$. Hence throughout the
paper matrix multiplication over different fields (one over the base
field and the other from the extended field) is allowed and computed
over the extended field.

There is a bijective mapping between $\F_p\hspace{-1.1mm}^{m\times
n}$ and $\F_{q}\hspace{-1.1mm}^{m}$ defined as follows:
\begin{itemize}
\item  For each $A\in\F_p\hspace{-1.1mm}^{m\times n}$, the folded version of $A$ is a vector $\mathbf{A}^f$
 in $\F_{q}\hspace{-1mm}^m$ given by $A \mathbf{a}^\text{T}$ where $\mathbf{a}=\{a_1,\ldots,a_n\}$ is a basis of the extension field $\F_{q}$ with respect to $\F_p$. Here we treat the $i^\text{th}$ row of $A$ as a single element in $\F_{q}$ to obtain the $i^\text{th}$ element of $A^f$. For instance let $A=\begin{bmatrix}1&0\\1&1\end{bmatrix}$ be a matrix in $\F_2\hspace{-1.1mm}^{2\times 2}$.
Then the operation of folding it into $\F_4\hspace{-1.1mm}^{2}$
gives $\mathbf{A}^f=\begin{bmatrix}(1,0)\\(1,1)\end{bmatrix}=
\begin{bmatrix} 2\\3\end{bmatrix}\in \F_4\hspace{-1.1mm}^{2}$ $($where $2\equiv x$ and $3\equiv x+1$ mod $(x^2+x+1)$~\cite{Algebra_Martin}$)$.

\item For each $\mathbf{B}\in\F_{q}\hspace{-1mm}^{m}$, the unfolded version of $\mathbf{B}$ is a matrix ${B}^u$ in $\F_p\hspace{-1.1mm}^{m\times n}$.
Here we treat the $i^\text{th}$ element of $\mathbf{B}$ as a row in
$\F_p\hspace{-1.1mm}^{1\times n}$ to obtain the $i^\text{th}$ row of
${B}^u$. For instance let
$\mathbf{B}=\begin{bmatrix}2\\3\end{bmatrix}$ be a vector in
$\F_4\hspace{-1.1mm}^{2}$. Then the operation of unfolding it into
$\F_2\hspace{-1.1mm}^{2\times 2}$ gives ${B}^u
=\begin{bmatrix}1&0\\1&1\end{bmatrix}$.
\end{itemize}

We can also extend these operations to include more general
scenarios. Specifically any matrix $A\in\F_p\hspace{-1.1mm}^{m\times
\ell n}$ can be written as a concatenation of matrices $A =
[A_1\ldots A_\ell]$, where $A_i\in F_p\hspace{-1.1mm}^{m\times n}$.
The folding operation is defined as follows: ${A}^f =
[\mathbf{A}_1^f\ldots \mathbf{A}_\ell^f]$. Similarly the unfolding
operation $u$ can be applied to a number of submatrices of a large
matrix, e.g., $[\mathbf{A}_1^f\ldots \mathbf{A}_\ell^f]^u=
[(\mathbf{A}_1^f)^u\ldots (\mathbf{A}_\ell^f)^u]=[A_1\ldots
A_\ell]$.

In the paper {\it double algebraic extensions} are considered. More
precisely let $\F_Q$ be an algebraic extension from $\F_q$, where
$Q=q^N=p^{nN}$ for any positive integer $N$.
Table~\ref{Tab:Field-Notation} summarize the notation of the fields
considered.

\begin{table}[h]
\caption{Summary of filed notations} 
\centering 
\begin{tabular}{c c cc}
\hline\hline 
Field&$\F_p$&$\F_q$&$\F_Q$
\\ [0.5ex]
\hline
Size &$p$&$q=p^n$&$Q=q^N $\\
\hline
\end{tabular}
\label{Tab:Field-Notation}
\end{table}

\noindent {\bf Note:} Of the three fields $\F_p$, $\F_{q}$ and
$\F_{Q}$ defined above, two or sometimes all three appear
simultaneously in the same equation. To avoid confusion, unless
otherwise specified, the superscript $f$ for folding is from  $\F_p$
to $\F_{q}$, and the superscript $u$ for unfolding is from $\F_{q}$
(or $\F_Q$) to $\F_p$.

\subsection{Row-space distance}

For any two matrices $B_1\in \F_p\hspace{-1.1mm}^{m_1\times n}$ and
$B_2\in \F_p\hspace{-1.1mm}^{m_2\times n}$ let $\mathcal {B}_1$ be
the subspace spanned by the rows of $B_1$ and $\mathcal {B}_2$ be
the subspace spanned by the rows of $B_2$. The {\it row-space
distance} of $B_1$ and $B_2$ is defined as $d_{S}(B_1,B_2)=dim(span
(\mathcal {B}_1\cup \mathcal {B}_2))-dim(span (\mathcal {B}_1\cap
\mathcal {B}_2))$. Row-space distance is a metric and satisfies the
triangle inequality~\cite{KotterSchk}.

If $m_1=m_2=m$, the following proposition is direct consequence of Corollary 3 in~\cite{RankMetricRandCodes}:
\begin{proposition}
$d_{S}(B_1,B_2)\leq 2 \mbox{ rank}(B_1-B_2)$.
\label{pro:rowdis-rankdis}
\end{proposition}
\subsection{Gabidulin codes}
\label{sec:Gabidulin}

Gabidulin in~\cite{Gabidulin_TheoryOfCodes_1985} introduced a class
of error correcting codes over $\F_p\hspace{-1.1mm}^{m\times n}$.
Let $\mathbf{X}\in\F_{q}^{R}$ be the information vector,
$G\in\F_{q}^{m\times R}$ be the generator matrix,
$(G\mathbf{X})^u\in\F_p ^{m\times n}$ be the transmitted matrix,
$Z\in\F_p ^{m\times n}$ be the error matrix, and $(G
\mathbf{X})^u+Z\in\F_p ^{m\times n}$ be the received matrix. Then
decoding is possible if and only if
rank$(Z)\leq\lfloor\frac{d}{2}\rfloor$, where $d=m-R+1$ is the
minimum distance of the code.

The work of~\cite{RankMetricRandCodes} utilizes the results of~\cite{Gabidulin_TheoryOfCodes_1985} to obtain network error-correcting codes with the following properties:
\begin{theorem}[Theorem 11 in~\cite{RankMetricRandCodes}]
Let $Z$ can be expressed as
$Z=\sum_{i\in[1,\tau]}\mathbf{L}_i\mathbf{E}_i$, such that:
\begin{itemize}
\item For each $i\in[1,\tau]$, $\mathbf{L}_i\in \F_p\hspace{-1.1mm}^{m\times 1}$ and $\mathbf{E}_i\in \F_p\hspace{-1.1mm}^{1\times n}$;

\item For each $i\in [1,\mu]$, $\mathbf{L}_i$ is known {\it a priori} by the
receiver;

\item For each $i\in [\mu+1,\mu+\delta]$, $\mathbf{E}_i$ is known {\it a priori} by the
receiver;

\item $2\tau-\mu-\delta\leq d-1$,
\end{itemize}
using Gabidulin codes the receiver can decode $\mathbf{X}$ with at
most $\mathcal O(mn)$ operations over $\F_{q}$.
\label{Th:GabiDecode}
\end{theorem}
Thus when $\mu=\delta=0$, Theorem~\ref{Th:GabiDecode} reduces to the
basic case where the receiver has no {\it priori} knowledge about
$Z$.

\section{Coherent network error-correcting codes}
\label{The_Coherent_Case}

Coherent here means the receiver $\mathcal{R}$ knows the linear
transforms from both $\mathcal{S}_1$ and $\mathcal{S}_2$, {\it
i.e.}, $\mathcal{R}$ knows $T_1$ and $T_2$ defined in equation
(\ref{eq:net_tran}). For instance, it is possible $T_1$ and $T_2$ to
be inferred by network communications before the adversary enters
the network and corrupts information. Alternatively, if centralized
designed network coding is used~\cite{Sid2003}, $T_1$ and $T_2$ is
assumed to be known by the receiver.

While the non-coherent codes we propose are more general than the
coherent codes, the description of the latter is simpler, and hence
we first describe them. Under the coherent assumption the goal of
the section is to construct a code attaining any rate-tuple
$(R_1,R_2)$ in the rate region for our communication scenario (see
section~\ref{The_Model} for details).

\subsection{Encoding}

Each source $\mathcal{S}_i$, $i\in[1,2]$, has information to deliver
to destination $\mathcal{R}$ and organizes this information into
batches of $R_i$ packets. Each packet is a concatenation of $
\ell=knN$ symbols from the finite field $\F_p$, where $n=R_1+2z$ and
$N=R_2+2z$ and $k$ is a code design parameter. For simplicity we
will analyze the transmission of a single batch of packets.

The way sources encode their information packets is through the use
of Gabidulin codes (see Section~\ref{sec:Gabidulin} for details).
More precisely the information of $\mathcal {S}_1$ is a matrix
${X}_1\in\F_{q} ^{R_1\times kN}$, where $\F_q$ is an algebraic
extension of $\F_p$ and $q = p^{n}$ (see
Section~\ref{sec:Field_extension} for details). Before transmission
${X}_1$ is multiplied with a generator matrix, $G_1\in\F_{q}
^{n\times R_1}$, creating $G_1 {X}_1\in\F_{q} ^{n\times (k N)}$
whose unfolded version $M_1=(G_1 {X}_1)^u$ is a matrix in
$\F_p\hspace{-1.1mm}^{n\times \ell}$ that is transmitted through the
network using the random linear network coding defined in
Section~\ref{subsec:random-linear-netcod}.

The information of $\mathcal {S}_2$ is a matrix
${X}_2\in\F_{Q}^{R_2\times k}$, where $\F_Q$ is an algebraic
extension of $\F_q$ and $Q = q^{N}=p^{nN}$. Before transmission
${X}_2$ is multiplied with a generator matrix,
$G_2\in\F_{Q}^{N\times R_2}$, creating $G_2 {X}_2\in\F_{Q}^{N\times
k }$ whose unfolded version $M_2=(G_2  {X}_2)^u$ over $\F_p$ is a
matrix in $\F_p\hspace{-1.1mm}^{N\times \ell}$ that is transmitted
through the network using the random linear network coding defined
in Section~\ref{subsec:random-linear-netcod}.

Both $G_1$ and $G_2$ are chosen as generator matrices for Gabidulin
codes and have the capability of correcting errors of rank at most
$z$ over $\F_p$ and $\F_{q}$ respectively.

\subsection{Decoding}
The packets reaching receiver $\mathcal{R}$ can be expressed as
\begin{eqnarray}
Y = T_1 M_1 + T_2 M_2 + E,
\label{eqn:Y_versus_X1_X2_E}
\end{eqnarray}
where $Y\in\F_p\hspace{-1.1mm}^{C\times  \ell}$ is the matrix formed
by the packets received by $\mathcal{R}$,
$T_1\in\F_p\hspace{-1.1mm}^{C\times n}$,
$T_2\in\F_p\hspace{-1.1mm}^{C\times N}$ are the linear transform
matrices from $\mathcal{S}_1$ and $\mathcal{S}_2$ to the receiver
$\mathcal{R}$, and $E\in \F_p\hspace{-1.1mm}^{C\times  \ell}$ is the
error matrix induced at the receiver. Note that rank($E$)$\leq z$
since the adversary can inject only $z$ error
packets~\cite{jaggi07resilient}.

Folding equation (\ref{eqn:Y_versus_X1_X2_E}) into $\F_{q}$ results
in:
\begin{eqnarray}
Y^f = [T_1G_1~~T_2] \left[
\begin{array}{c}
{X}_1\\
{M}_2^f
\end{array}
\right]+{E}^f \label{eqn:Y_versus_X1_X2_E_embedded1},
\end{eqnarray}
where $E^f$ has rank at most equal to $z$ according to
Lemma~\ref{lemma:RankBoundE}.
\begin{lemma}
\label{lemma:RankBoundE}
Folding a matrix does not increase its rank.
\end{lemma}
\noindent{\it Proof}: Let  matrix $H\in\F_p\hspace{-1.1mm}^{m\times
k n}$ has rank$(H)=r$. Thus $H=W Z$, where
$Z\in\F_p\hspace{-1mm}^{r\times k n}$ is of full row rank and
$W\in\F_p\hspace{-1mm}^{m\times r}$ is of full column rank. After
the folding operation $H$ becomes ${H}^f = W {Z}^f$ and therefore
rank$({H}^f)\leq r$. \hfill$\Box$

Let $D=[\hspace{1mm}T_1G_1\hspace{1mm}~T_2\hspace{1mm}]$. Since
$R_1+N=R_1+R_2+2z=C$ (see Section~\ref{The_Model} for details), $D$
is a
$C\times C$ square matrix.%
\begin{lemma}
\label{lemma:Invertible_D} Matrix $D\in\F_{q}^{C\times C}$ is
invertible with probability at least $1-\NetSize/p$.
\end{lemma}
\noindent{\it Proof}: Let $\mathcal X$  be the set of random
variables over $\F_p$ comprised of the local coding coefficients
used in the random linear network code. Thus the determinant of $D$
is a polynomial ${\bf f}(\mathcal X)$ over $\F_{q}$ of degree at
most $\NetSize$ (see Theorem 1 in~\cite{RandCode0} for details).
Since the variables $\mathcal X$ in ${\bf f}(\mathcal X)$ are
evaluated over $\F_p$, ${\bf f}(\mathcal X)$ is equivalent to a
vector of polynomials $(f_1(\mathcal X),f_2(\mathcal
X),\ldots,f_{n}(\mathcal X))$, where $f_i(\mathcal X)\in
\F_p[\mathcal X]$ is a polynomial over $\F_p$ with variables in
$\mathcal X$. Note that $f_i(\mathcal X)$ also has degree no more
than $\NetSize$ for each $i\in [1,n]$. Thus once we prove that there
exists an evaluation of $\mathcal X$ such that ${\bf f}$ is a
nonzero vector over $\F_p$, we can show $D$ is invertible with
probability at least $1-\NetSize/p$ by Schwartz-Zippel
lemma~\cite{CCBook}.

Since $R_1+N= C$ (see Section~\ref{The_Model} for details) and
$R_1\leq C_1$ and $N\leq C_2$, there exist $R_1+N$
edge-disjoint-paths:
$\DisPath^1_1,\DisPath^1_2,\ldots,\DisPath^1_{R_1}$ from $s_1$ to
$r$ and $\DisPath^2_1,\DisPath^2_2,\ldots,\DisPath^2_{N}$ from $s_2$
to $r$. The variables in $\mathcal X$ are evaluated in the following
manner:

1). Let $O$ be the zero matrix in $F_{q}^{n \times N}$. We choose
the variables in $\mathcal X$ so that the $R_1$ independent rows of
$[G_1, O] \in \F_{q}^{n \times C}$ correspond to routing information
from $s_1$ to $R$ via $\DisPath_1^1, \ldots, \DisPath_{R_1}^1$.

2). Let $\{{\bf u}_{R_1+1},{\bf u}_{R_1+2},\ldots,{\bf u}_{C}\}$ be
$N$ distinct rows of the identity matrix in $\F_{q} ^{C\times C}$
such that for each $i\in[1,N]$, ${\bf u}_{R_1+i}$ has the element
$1$ located at position $R_1+i$. Then these $N$ vectors correspond
to routing information from $s_2$ to $r$ via
$\DisPath^2_1,\DisPath^2_2,\ldots,\DisPath^2_{N}$.

Under such evaluations of the variables in $\mathcal X$, matrix $D$
equals $\begin{bmatrix} G_1'&O\\O &I_{N}\end{bmatrix}$, where
$G_1'\in \F_{q} ^{R_1\times R_1}$ consists  of the $R_1$ independent
rows of $G_1$. Hence ${\bf f}$ is non-zero. Using the
Schwartz-Zippel Lemma ${\bf f}\neq 0$ and thus $D$ is invertible
with probability at least $1-\NetSize/p$ over the choices of
$\mathcal X$. \hfill$\Box$

Hence, by multiplying
Equation~(\ref{eqn:Y_versus_X1_X2_E_embedded1}) by $D^{-1}$ the
receiver gets  $D^{-1} {Y}^f =
\begin{bmatrix}
{X}_1\\ {M}_2^f
\end{bmatrix}
+ D^{-1}  {E}^f$. The last $N=R_2+2z$ rows of $D^{-1} {Y}^f$ are
$(D^{-1} {Y}^f)_d =
 {M}_2^f + (D^{-1} {E}^f)_d$, where the subscript $d$ stands
for the last $N$ rows of each matrix.

\noindent {\bf Note:} To show why $\mathcal{S}_2$ uses a generator
matrix $G_2$ over a double-extended field $\F_{Q}=
\F_{q^{N}}=\F_{p^{nN}}$, consider what happens if instead it uses
$\F_Q=\F_q$. In this case the matrix $ {M}_2^f+(D^{-1} {E}^f)_d$ is
indeed of the form required for successful decoding of Gabidulin
codes as long as $(D^{-1} {E}^f)_d^u$ has rank less than $z$ over
$\F_p$. But this is not generally the case since $D^{-1}$ belongs to
$\F_{q}$ but not $\F_p$. Therefore although $ {E}^f$ and
consequently $D^{-1} {E}^f$ have rank less than $z$ over $\F_{q}$,
the rank of $(D^{-1} {E}^f)_d^u$ might increase over $\F_p$.

If source $\mathcal {S}_2$ uses a generator matrix $G_2$ defined
over $\F_{Q}=\F_{q^{N}}$ that is able to correct rank $z$ errors
over $\F_{q}$, we can prove the main result in this section as
follows.

\begin{theorem}
A coherent receiver $\mathcal{R}$ can efficiently decode both $X_1$
and $X_2$ correctly with probability at least $1-2\NetSize/p$.
\label{Th:Coherent}
\end{theorem}

\noindent{\it Proof}: First, according to
Lemma~\ref{lemma:Invertible_D} matrix $D$ is invertible with
probability at least $1-\NetSize/p$. Since $G_2$ is able to correct
rank $z$ errors over $\F_{q}$, using $(D^{-1}  Y^f)_d = M_2^f +
(D^{-1}E^f)_d$, $\mathcal{R}$ can execute the Gabidulin decoding
algorithm and get $X_2$.

Second, once $X_2$ is known $T_2M_2$ is subtracted from $Y$ to
result in $T_1M_1+E$. Since $T_1$ is left invertible with
probability at least $1-\NetSize/p$ (by~\cite{RandCode2}), $\mathcal
R$ can multiply $T_1M_1+E$ with the left inverse of $T_1$ giving
$M_1+T_1^{-1}E$. Since rank$(T_1^{-1}E)\leq z$ over base field
$\F_p$, the execution of the Gabidulin decoding algorithm results in
$X_1$. In the end the overall probability of correct decoding is at
least $1-2\NetSize/p$. \hfill$\Box$
\subsection{Complexity discussion}
Since the computational complexity of the coherent network
error-correcting codes here is that same as those of the
non-coherent codes shown later, we delay the discussion until
Section~\ref{The_Complexity_Discussion}.

\section{Non-Coherent Error correction}
\label{The_Non_Coherent_Case} In the non-coherent case it is assumed
that receiver $\mathcal{R}$ does not know the network transform
matrices $T_1$ and $T_2$ of the two sources prior to communication
in the presence of the adversary. Assuming a non-coherent receiver
the goal of this section is to construct codes attaining any
rate-tuple $(R_1,R_2)$ in the rate region for our communication
scenario (see section~\ref{The_Model} for details).

\subsection{Encoding}

In the scenario where the receiver $\mathcal{R}$ does not know $T_1$
and $T_2$ {\it a priori} the two sources append  headers on their
transmitted packets to convey information about $T_1$ and $T_2$ to
the receiver. Thus source $\mathcal{S}_1$ constructs message matrix
$[I_{n}\hspace{2.5mm}O\hspace{2.5mm}M_1]$ with the zero matrix $O$
having dimensions $n\times N$, and source $\mathcal{S}_2$ constructs
a message matrix $[O\hspace{2.5mm}I_{N}\hspace{2.5mm}M_2]$ with the
zero matrix $O$ having dimension $N\times n$. The identity and zero
matrices have elements from $\F_p$ and the $M_1$, $M_2$ matrices in
$\F_p^{C\times \ell}$ have the same definitions as in
Section~\ref{The_Coherent_Case}-A.

\subsection{Decoding}

The two message matrices are transmitted to the receiver
$\mathcal{R}$ through the network with the use of random linear
network code and therefore the receiver gets:
\begin{eqnarray}
Y = [Y_1\hspace{2.5mm}Y_2\hspace{2.5mm}Y_3] =
[T_1\hspace{2.5mm}T_2\hspace{2.5mm}A]+E, \label{eqn:Y_Y1_Y2_Y3}
\end{eqnarray}
where $A = T_1M_1 + T_2M_2\in \F_p\hspace{-1.1mm}^{C\times  \ell}$
and $E\in\F_p\hspace{-1.1mm}^{C\times(n+N+ m)}$ has rank no more
than $z$ over field $\F_p$. Let $E=\begin{bmatrix} E_1&E_2
&E_3\end{bmatrix}$, where $E_1\in \F_p\hspace{-1.1mm}^{C\times n}$
and    $E_2\in \F_p\hspace{-1.1mm}^{C\times N}$ and $E_3\in
\F_p\hspace{-1.1mm}^{C\times  \ell}$. As in the decoding scheme in
Section~\ref{The_Coherent_Case} the receiver $\mathcal R$ first
decodes $X_2$ and then $X_1$.

{\bf Stage 1: Decoding $X_2$:}

Let $Y_a=\begin{bmatrix}Y_1G_1& Y_2 & Y_3^f\end{bmatrix}$ be a
matrix in $\F_{q} ^{C\times(R_1+N+kN)}$. To be precise:
\begin{eqnarray}Y_a
& =&
 \begin{bmatrix}
  T_1G_1&T_2&A^f
 \end{bmatrix}
 +
 \begin{bmatrix}
  E_1G_1 & E_2 & E_3^f
 \end{bmatrix}.\label{eq:Y_a}
\end{eqnarray}
 Receiver $\mathcal{R}$ uses invertible row operations over
$\F_{q}$ to transform $Y_a$ into a row-reduced echelon matrix
$\begin{bmatrix} T_{RRE}&M_{RRE}\end{bmatrix}$ that has the same row
space as $Y_a$, where $T_{RRE}$ has $C=R_1+N$ columns and $M_{RRE}$
has $kN$ columns. Then the following propositions are from the
results\footnote{1) is from Proposition 7, 2) from Theorem 9, and 3)
from Proposition 10 in~\cite{RankMetricRandCodes}.} proved
in~\cite{RankMetricRandCodes}:
\begin{proposition}
\begin{enumerate}
\item \label{subPro:RRE}  The matrix $\begin{bmatrix} T_{RRE}&M_{RRE}\end{bmatrix}$ takes
the form
 $\begin{bmatrix} T_{RRE}&M_{RRE}\end{bmatrix}=\begin{bmatrix}I_{C}+\hat{L}{U}_\mu^T&r\\O&\hat{E}\end{bmatrix}$,
where ${U}_\mu\in \F_{q} ^{C\times \mu}$ comprises of $\mu$ distinct
columns of the $C\times C$ identity matrix such that ${U}_\mu^Tr=0$
and ${U}_\mu^T\hat L=-I_{\mu}$. In particular, $\hat{L}$ in $\F_{q}
^{C\times \mu}$ is the ``error-location matrix", $r$ in $\F_{q}
^{C\times kN}$ is the ``message matrix", and $\hat E$ in
$\F_{q}\hspace{-1.4mm}^{\delta \times kN}$ is the ``known error
value" (and its rank is denoted $\delta$).

\item \label{subPro:Row-dis} Let  $X=\begin{bmatrix} X_1 \\
M_2^f\end{bmatrix}$ and $e=r-X$ and $\tau=\text{rank}\begin{bmatrix} \hat{L} & e\\
0 &\hat{E} \end{bmatrix}$. Then $2\tau-\mu-\delta$ is no more than
$d_S(\begin{bmatrix} T_{RRE} & M_{RRE}
\end{bmatrix},\begin{bmatrix} I_{C}&X\end{bmatrix})$, {\it i.e.}, the row-space distance between $\begin{bmatrix}
T_{RRE} & M_{RRE}
\end{bmatrix}$ and $\begin{bmatrix} I_{C}&X\end{bmatrix}$.

\item \label{subPro:errorform} There exist $\tau$ column vectors
$\mathbf L_1,\mathbf L_2,\ldots,\mathbf L_\tau\in\F_{q} ^{C}$ and
$\tau$ row vectors $\mathbf E_{1},\mathbf E_{2},\ldots,\mathbf
E_{\tau}\in\F_{q} ^{1\times kN}$ such that $e=\sum_{i\in [1,\tau]}
\mathbf L_i \mathbf E_{i}$. In particular, $\mathbf L_1,\mathbf
L_2,\ldots,\mathbf L_{\mu}$ are the columns of $\hat{L}$, and
$\mathbf E_{\mu+1},\mathbf E_{\mu+2},\ldots,\mathbf E_{\mu+\delta}$
are the rows of $\hat{E}$.
\end{enumerate}
\label{pro:RRE}
\end{proposition}

Recall that the subscripte $d$ stands for the last $N$ rows of any
matrix/vector. Then we show the following for our scheme.
\begin{lemma}
1) Matrix $\hspace{1mm} e_d=r_d-M_2^f$ can be expressed as
$e_d=\sum_{i\in 1,2,\ldots,\tau}(\mathbf L_i)_d \mathbf E_i$, where
$(\mathbf L_1)_d,(\mathbf L_2)_d,\ldots,(\mathbf L_{\mu})_d$ are the
columns of $\hat L_d$ and $\mathbf E_{\mu+1},\mathbf
E_{\mu+2},\ldots,\mathbf E_{\mu+\delta}$ are the rows of $\hat{E}$.

$2)\hspace{1mm}$With probability at least $1-\NetSize/p$,
\begin{eqnarray*}
2\tau-\mu-\delta\leq 2z
\end{eqnarray*}
\label{Le:CorrectX2}
\end{lemma}
\vspace{-7mm}

\noindent{\it Proof}: $1)\hspace{1mm}$It is a direct corollary from
Proposition~\ref{pro:RRE}.3.

$2)\hspace{1mm}$Using Proposition~\ref{pro:RRE}.\ref{subPro:Row-dis}
it suffices to prove with probability at least $1-\NetSize/p$,
$d_S(\begin{bmatrix} T_{RRE} & M_{RRE}\end{bmatrix}, \begin{bmatrix}
I_{C}&X\end{bmatrix})\leq 2z$.

As shown in the proof of Lemma~\ref{lemma:RankBoundE}, the columns
of $E_3^f$ are in the column space of $E_3$ (and then of $E$) over
$\F_{q}$. Thus $\begin{bmatrix}E_1&E_2&E_3^f\end{bmatrix}$ and
therefore $\begin{bmatrix}E_1G_1&E_2&E_3^f\end{bmatrix}$ has rank at
most equal to $z$ over $\F_{q}$. Using
Proposition~\ref{pro:rowdis-rankdis} and Equation~(\ref{eq:Y_a}),
$d_S(Y_a,\begin{bmatrix}T_1G_1&T_2&A^f\end {bmatrix})$ is no more
than $2z$.

Since $d_S(\begin{bmatrix} T_{RRE} & M_{RRE}\end{bmatrix},Y_a)=0$,
we have $d_S(\begin{bmatrix} T_{RRE} &
M_{RRE}\end{bmatrix},\begin{bmatrix}T_1G_1&T_2&A^f\end
{bmatrix})\leq 2z$.

Using Lemma~\ref{lemma:Invertible_D}, matrix $D$ is invertible with
probability at least $1-\NetSize/p$, so $\begin{bmatrix} I_{C}&X\end
{bmatrix}$ has zero row-space distance from $\begin{bmatrix}
D&DX\end {bmatrix}=\begin{bmatrix}T_1G_1&T_2&A^f\end {bmatrix}$.
Thus $d_S(\begin{bmatrix} T_{RRE} &
M_{RRE}\end{bmatrix},\begin{bmatrix} I_{C}&X\end{bmatrix})\leq 2z$.
\hfill$\Box$

In the end combining Lemma~\ref{Le:CorrectX2} and
Theorem~\ref{Th:GabiDecode} the receiver $\mathcal{R}$ can take
$(\hat L_d, \hat E, r)$ as the input for the Gabidulin decoding
algorithm and decode $X_2$ correctly.

 {\bf Stage 2: Decoding
${X_1}$:}

From equation (\ref{eqn:Y_Y1_Y2_Y3}) the receiver $\mathcal{R}$ gets
$Y=\begin{bmatrix}T_1+E_1&T_2+E_2&A+E_3\end{bmatrix}$. The receiver
$\mathcal{R}$ computes $(T_2+E_2)M_2$  and then it subtracts matrix
$\begin{bmatrix}O&(T_2+E_2)& (T_2+E_2)M_2\end{bmatrix}$ from $Y$.
The resulting matrix has $N$ zero columns in the middle (column
$n+1$ to column $n+N$). Disregarding these we get:
\begin{eqnarray*}
Y' = \begin{bmatrix}T_1&T_1 M_1\end{bmatrix}+
\begin{bmatrix}E_1&E_3-E_2 M_2\end{bmatrix}.
\end{eqnarray*}

The new error matrix $E' =
\begin{bmatrix}E_1&E_3-E_2 M_2\end{bmatrix}$ has rank at
most $z$ over $\F_p$ since the columns of $E'$ are simply linear
combinations of columns of $E$ whose rank is at most $z$. Therefore
the problem degenerates into a single source problem and receiver
$\mathcal{R}$ can decode $X_1$ with probability at least
$1-\NetSize/p$ by following the approach in
\cite{RankMetricRandCodes}.

Summarizing the above decoding scheme for $X_1$ and $X_2$, we have
the main result in the section.
\begin{theorem}
A non-coherent receiver $\mathcal{R}$ can efficiently decode both
$X_1$ and $X_2$ correctly with probability at least $1-2\NetSize/p$.
\label{Th:NonCoherent}
\end{theorem}
\vspace{-7mm}
\subsection{Complexity discussion} \label{The_Complexity_Discussion}
The paper consider the technique of double field-extension to design
double-access network  codes robust against network errors. The
technique has not been considered before in the literature, and
makes achieving the rate-region proved in~\cite{Svit_rate_regions}
computationally tractable.

For both coherent and non-coherent cases the computational
complexity of Gabidulin encoding and decoding of two source messages
is dominated by the decoding of $X_2$, which requires $\mathcal O(C
\ell nN \log(n N))$ operations over $\F_p$
(see~\cite{RankMetricRandCodes}).

To generalize our technique to more sources, consider a network with
$\SourceNum$ sources $\mathcal {S}_1,\mathcal {S}_2,\ldots,\mathcal
{S}_\SourceNum$. Let $R_i$ be the rate of $\mathcal {S}_i$ and
$n_i=R_i+2z$ for each $i\in[1,\SourceNum]$. A straightforward
generalization uses the multiple field-extension technique so that
$\mathcal{S}_i$ uses the generator matrix over finite field of size
$p^{n_1 n_2\ldots n_i}$. In the end the packet length must be at
least $n_g=n_1 n_2\ldots n_\SourceNum$, resulting in a decoding
complexity $\mathcal O(C n_g^2 \log(n_g))$ increasing exponentially
in the number of sources $\SourceNum$. Thus the multiple
field-extension technique works in polynomial time only for a fixed
number of sources.

\bibliography{NWC-abbr}

\begin{thebibliography}{10}
\providecommand{\url}[1]{#1}
\csname url@rmstyle\endcsname
\providecommand{\newblock}{\relax}
\providecommand{\bibinfo}[2]{#2}
\providecommand\BIBentrySTDinterwordspacing{\spaceskip=0pt\relax}
\providecommand\BIBentryALTinterwordstretchfactor{4}
\providecommand\BIBentryALTinterwordspacing{\spaceskip=\fontdimen2\font plus
\BIBentryALTinterwordstretchfactor\fontdimen3\font minus
  \fontdimen4\font\relax}
\providecommand\BIBforeignlanguage[2]{{%
\expandafter\ifx\csname l@#1\endcsname\relax
\typeout{** WARNING: IEEEtran.bst: No hyphenation pattern has been}%
\typeout{** loaded for the language `#1'. Using the pattern for}%
\typeout{** the default language instead.}%
\else
\language=\csname l@#1\endcsname
\fi
#2}}

\bibitem{ahlswede00network}
R.~Ahlswede, N.~Cai, S.-Y. Li, and R.~Yeung, ``Network information flow,''
  \emph{IEEE Transactions on Information Theory}, vol.~46, no.~4, pp.
  1204--1216, July 2000.

\bibitem{RaymondNECC2002}
N.~Cai and R.~W. Yeung, ``Network coding and error correction,'' in \emph{Proc.
  of 2002 IEEE Information Theory Workshop (ITW)}, 2002.

\bibitem{jaggi07resilient}
S.~Jaggi, M.~Langberg, S.~Katti, T.~Ho, D.~Katabi, and M.~M\'edard, ``Resilient
  network coding in the presence of byzantine adversaries,'' in \emph{Proc.
  IEEE INFOCOM 2007}, 2007.

\bibitem{RankMetricRandCodes}
D.~Silva, F.~Kschischang, and R.~Koetter, ``A rank-metric approach to error
  control in random network coding,'' \emph{IEEE Transactions on Information
  Theory}, vol.~54, no.~9, pp. 3951--3967, Sept. 2008.

\bibitem{gkantsidis06:cooperative}
C.~Gkantsidis and P.~Rodriguez, ``Cooperative security for network coding file
  distribution,'' in \emph{Proc. of the 25th IEEE INFOCOM}, 2006.

\bibitem{charles06signaturesfor}
D.~Charles, K.~Jain, and K.~Lauter, ``Signatures for network coding,'' in
  \emph{Proc. of the Fortieth Annual Conference on Information Sciences and
  Systems}, 2006.

\bibitem{Svit_rate_regions}
S.~Vyetrenko, T.~Ho, M.~Effros, J.~Kliewer, and E.~Erez, ``Rate regions for
  coherent and noncoherent multisource network error correction,'' in
  \emph{Proc. of ISIT 2009}, 2009.

\bibitem{RandCode0}
T.~Ho, M.~Medard, J.~Shi, M.~Effros, and D.~R. Karger, ``On randomized network
  coding,'' in \emph{Proc. of Allerton 2003}, 2003.

\bibitem{Algebra_Martin}
M.~Artin, \emph{Algebra}.\hskip 1em plus 0.5em minus 0.4em\relax New Jersey:
  Prentice Hall, 1991.

\bibitem{KotterSchk}
R.~K\"otter and F.~R. Kschischang, ``Coding for errors and erasures in random
  network coding,'' \emph{IEEE Transactions on Information Theory}, vol.~54,
  no.~8, pp. 3579--3591, Aug. 2008.

\bibitem{Gabidulin_TheoryOfCodes_1985}
E.~M. Gabidulin, ``Theory of codes with maximum rank distance,'' \emph{Probl.
  Peredachi Inf.}, vol.~21, no.~1, pp. 3--16, 1985.

\bibitem{Sid2003}
S.~Jaggi, P.~Sanders, P.~A. Chou, M.~Effros, S.~Egner, K.~Jain, and
  L.~Tolhuizen, ``Polynomial time algorithms for multicast network code
  construction,'' \emph{IEEE Transactions on Information Theory}, vol.~51,
  no.~6, pp. 1973--1982, 2003.

\bibitem{CCBook}
M.~Agrawa and S.~Biswas, ``Primality and identity testing via chinese
  remaindering,'' \emph{Journal of the ACM}, 2003.

\bibitem{RandCode2}
P.~A. Chou, Y.~Wu, and K.~Jain, ``Practical network coding,'' in \emph{Proc. of
  Allerton 2003}, 2003.

\end{thebibliography}
\bibliographystyle{IEEEtran}
\end{document}